\documentstyle[twocolumn,eclepsf]{jpsj}
\newcommand{\simle}
{\raisebox{-0.75ex}[-1.5ex]{$\;\stackrel{<}{\sim}\;$}}
\newcommand{\simge}
{\raisebox{-0.75ex}[-1.5ex]{$\;\stackrel{>}{\sim}\;$}}

\def\e{{\epsilon}}
\def\k{{ {\bf k} }}
\def\p{{ {\bf p} }}
\def\q{{ {\bf q} }}
\def\Q{{ {\bf Q} }}
\def\w{{\omega}}
\def\a{{\alpha}}

\def\l{{\lambda}}

\def\X{\times}

\def\TMTSFPF{ (TMTSF)$_2$PF$_6$ }
\def\TMTTFPF{ (TMTTF)$_2$PF$_6$ }

\def\runtitle{Effects of Spin Fluctuations 
in  Quasi-One-Dimensional Organic Superconductors}

\def\runauthor{Hiori {\sc Kino} and  Hiroshi {\sc Kontani}}

\title{Effects of Spin Fluctuations \\
in  Quasi-One-Dimensional Organic Superconductors}

\author{ Hiori {\sc Kino} and Hiroshi {\sc Kontani}$^1$ }

\inst{ Joint Research Center for Atom Technology, Tsukuba, Ibaraki, 305-0046,\\
$^1$Institute for Solid State Physics, University of Tokyo, 7-22-1 Roppongi, Minato-ku, Tokyo 106-8666.}

\recdate{February 23, 1999}

\abst{
We study the electronic states 
        of  quasi-one-dimensional organic conductors
        using the single band Hubbard model at half-filling.
We treat the effects of  the on-site Coulomb interaction 
        by the fluctuation-exchange (FLEX) method, 
 and calculate the phase diagram and physical properties.
The calculated pressure dependence of the N\'eel  temperature
 coincides well with the experimental one.
We also show that a pseudogap is 
        formed in the density of states near the chemical potential 
        and that $d$-wave superconductivity appears 
         next to the antiferromagnetic state.
Moreover the NMR relaxation rate increases on cooling 
in the low-temperature region
}

\kword{
quasi-one-dimensional organic conductors, Hubbard model, spin fluctuations, TMTSF, TMTTF, FLEX approximation, superconductivity, antiferromagnetism 
}

\begin{document}

\sloppy
        \maketitle

It is known that 
the quasi-one-dimensional organic superconductors (TMTSF)$_2$X and 
 similar compounds
have very rich varieties of phases.\cite{Ishiguro1989,Moser1998}
For example,  \TMTTFPF is
 in the spin-Peierls phase  at ambient pressure.
The antiferromagnetic insulating phase is stabilized 
with applied pressure larger than about 10 kbar.
The antiferromagnetic transition temperature, $T_{\rm N}$,
becomes maximum ($\sim 20$ K)  around 15 kbar.
\TMTSFPF is in the antiferromagnetic insulating phase 
  at ambient pressure.
The antiferromagnetism disappears with increasing pressure
 to give rise to superconductivity.
The maximum superconducting transition temperature is about 1 K.
The superconducting phase is next to the antiferromagnetic one and 
it indicates that the spin fluctuation is closely connected with the emergence of the superconductivity.
It is also very interesting that the pseudogap was
        observed in the density of states (DOS)  near the chemical potential 
        in XPS experiments.\cite{Dardel1993,Zwick1997}

In the previous theoretical studies, 
the Hubbard model was mainly studied
 as the simplest model in these systems.
The SDW  transition temperature and phase boundary between the SDW and the superconducting phases were discussed within the mean field, or the RPA treatment.\cite{Yamaji1983,Hasegawa1986}
Shimahara discussed the spin fluctuations as
 an origin of the superconductivity.\cite{Shimahara1989}
This work and the QMC study indicated 
that the superconductivity was realized and 
the superconducting order parameter had line-nodes on the Fermi surface.\cite{Kuroki1998}
Dimensional crossover was discussed based on the 
renormalization group theory.\cite{Kishine1998}


The aim of this letter is to discuss 
  the origin of pseudogap behavior detected in the XPS experiments,
  and the origin of superconductivity on the basis of the Fermi liquid theory, 
and to understand the phase diagrams of TMTSF and TMTTF salts.
We show that these physical behaviors can be explained naturally
 in terms of  spin-fluctuation theory.
This is the first study on the quasi-one-dimensional organic conductors
where the effects of spin fluctuations are taken into account 
 in a self-consistent way
 and where the antiferromagnetism and the superconductivity 
   are treated on the same footing.

These systems have a half hole per TMTSF/TMTTF molecule in the HOMO level,
 and   the one-dimensional chain is connected weakly. \cite{Grant1983,Jacobsen1983}
The dimerization, though it is weak,  exists and the antibonding band
which crosses the chemical potential  is half-filled.\cite{Penc1994}
Therefore we consider a single-band Hubbard model 
whose  band dispersion is given as
\begin{equation}
   \epsilon(k_x,k_y) = 2 t_0 \cos a k_x + 2 t_1  \cos b k_y + 2 t_2 \cos 2 b k_y
 \mbox{,}
\label{eq-half}
\end{equation}
where $|t_0|\gg |t_1| \gg |t_2|$ and 
$a$ and $b$ are the effective lattice constants.
It was known that the $\cos 2 b k_y$ term was necessary
 to  appropriately reproduce the shape of  the Fermi surface.\cite{Hasegawa1986}
The physical role of this term is 
that it  breaks the nesting condition, in other words, 
that it introduces frustration.
It was also known that 
         $t_2$ was  proportional to $t_1^2/t_0$,
        when this  term  was derived 
        from the quarter-filled Hubbard model.
We study the single-band Hubbard model at half-filling 
 and assume the screening of long range Coulomb interactions 
        for simplicity.
This is justified 
  because the antiferromagnetic fluctuations play the essential roles.

In this study, we put $t_0=-1$, $t_2 = 0.8 t_1^2/t_0$ and 
the on-site Coulomb interaction $U=3.0$, and vary $t_1 (<0)$ as a parameter. 
Here $|t_2|$ is rather large compared to the model derived from the quarter-filled Hubbard model.\cite{Hasegawa1986}
Such a choice is mainly due to the convenience of the numerical calculation because we use a finite $\k$-mesh.
We implicitly use 128$\X$128 $\k$-mesh and 256 Matsubara frequencies
 in this study.

To tackle the present model
we employ the fluctuation exchange (FLEX) method, 
which is a kind of self-consistent perturbation theory 
with respect to $U$.
The FLEX method has been considered to be advantageous
         for systems with large spin fluctuations.
It was known that this method gave the Green function 
which agreed with the QMC one 
        for the square-lattice Hubbard model 
        with moderate $U$. \cite{Bickers1991}
This method was applied to the study of high-$T_{\rm c}$ cuprates, superconducting ladder compounds and $\kappa$-(BEDT-TTF) salts and so on,\cite{Bickers1989,Kontani1998,Kino1998}
and predicted the possible $d$-wave superconductivity.
The QMC study also supported this result.\cite{Kuroki1998,Kuroki1998highTc}

Here, we explain the FLEX method.
The Dyson equation is written as
\begin{eqnarray}
\left\{ {G}(\k,\e_n) \right\}^{-1}
= \left\{ {G}^0(\k,\e_n) \right\}^{-1} - {\Sigma}(\k,\e_n),
 \label{eqn:Dyson}
\end{eqnarray}
where ${G}^0(\k,\e)$ is the unperturbed Green function.
The self-energy at temperature $T$ is given by
\begin{eqnarray}
& &\Sigma(\k,\e_n) 
 = T\sum_{\q,l} G(\k-\q,\e_n-\w_l)\cdot U^2    \nonumber \\
& &\ \times \left( \frac32 {\chi}^{(-)}(\q,\w_l) 
  +\frac12 {\chi}^{(+)}(\q,\w_l) 
  - {\chi}^0(\q,\w_l) \right) \mbox{,}
     \label{eqn:self} \\
& &{\chi}^{(\pm)}(\q,\w_l)
 = {\chi}^0 (\q,\w_l)  \cdot \left\{ { 1} \pm 
  U{\chi}^0(\q,\w_l) \right\}^{-1} \mbox{,} 
     \label{eqn:chi} \\
& &\chi^0(\q,\w_l)
 = -T\sum_{\k, n} G(\q+\k,\w_l+\e_n) G(\k,\e_n) \mbox{,}
     \label{eqn:chi0}
\end{eqnarray}
where $\e_n= (2n+1)\pi T$ and $\w_l= 2l\pi T$.
Equations (\ref{eqn:Dyson})-(\ref{eqn:chi0}) are calculated
together with the equation 
for the chemical potential $\mu$ given by
\(
N  =  2 T
\sum_{\p,n}  G(\p,\e_n) e^{i \e_n 0^+ } \mbox{,}
\)
where $N=1$ in this system.

To determine the magnetic transition temperature $T_{\rm N}$, 
we calculate the Stoner factor 
without vertex corrections, 
\(
\a_{\rm S}= \max_{\k}\left\{ \ U\cdot \chi^{0}(\k,\w\!=\!0)\  \right\}
\).
The antiferromagnetic critical points are determined
by the Stoner criterion, $\a_{\rm S}= 1$.

We also determine the superconducting transition temperature $T_{\rm c}$
by solving the linearized Eliashberg equation with respect to 
the singlet-pairing order parameter,
$\phi(-\k,\e_n)= + \phi(\k,\e_n)$,
\begin{eqnarray}
& &\lambda \cdot \phi(\k,\e_n)= -T\sum_{\q, m}
 V(\k-\q,\e_n-\e_m)  \nonumber \\
& & \ \ \times G(\q,\e_m) G(-\q,-\e_m)
 \cdot \phi(\q,\e_m), \label{eqn:lambda}\\
& &{V}(\k,\w_l)=  \frac32 U^2 {\chi}^{(-)}(\k,\w_l)
 - \frac12 U^2 {\chi}^{(+)}(\k,\w_l) + U ,
 \label{eqn:V}
\end{eqnarray}
where $T_{\rm c}$ is given by the  condition $\lambda=1$.

The theories  by Mermin and  Wagner, 
 and Hohenberg 
 prohibit finite  $T_{\rm N}$ and $T_{\rm c}$\ 
in two dimensions.
It was well known that $\a_{\rm S}$ 
satisfies this condition, 
because FLEX treats the spin fluctuations self-consistently.\cite{high-Tc-Dahm}
We estimate $T_{\rm N}$\ by the condition $\a_{\rm S}= \a_{\rm N}$,
where  $(1-\a_{\rm N})^{-1} \sim O(100)$.
This implies that the weak magnetic coupling between layers $J_\perp$
 makes the system  ordered three-dimensionally.
On the other hand, $\l=1$ is fulfilled at  finite $T_{\rm c}$\ 
 using eq.\,(\ref{eqn:lambda}).
However the $T_{\rm c}$ given by the Eliashberg equation is reliable in many cases.


First, let us examine the phase diagram on the plane  of $|t_1|$ and $T$.
Each line corresponds to $T_{\rm N}$ for different  $\a_{\rm N}$.
With increasing $|t_1|$ ($\simle 0.3$)
$T_{\rm N}$  increases gradually 
 and  has a broad peak ($T_{\rm N}\sim 0.4$) around $|t_1|= 0.3$.
On the other hand, $T_{\rm N}$ decreases rather rapidly 
with increasing  $|t_1|$ ($\simge 0.4$).
This behavior is consistent with experiments 
 if one regards $|t_1|$ as applied pressure.
We do not have a  plot for $|t_1|<0.1$,
         where $\k$-mesh and Matsubara frequencies are 
 insufficient to determine $T_{\rm N}$ numerically.

An inset of Fig.~\ref{phasediagram} shows the phase diagram at $U=1.8$
where $\chi(\q,\w)$ is calculated using $G^0(\k,\w)$ and 
the N\'eel temperature is estimated by the Stoner criterion, 
$\a_{\rm N}=1$.
The value of $U$ is chosen
in order to make $T_{\rm N}$ drop at $|t_1|\sim 0.4$.
In this case $T_{\rm N}$ does not descend with decreasing $|t_1|$.  
Therefore the decrease of $T_{\rm N}$ for small $|t_1|$  in Fig.~\ref{phasediagram} is due to the self-consistency.
We note that 
the estimated $T_{\rm N}$ using $G^0(\k,\w)$ is about 10 times larger than that using self-consistent $G(\k,\w)$, at $t_1=-0.3$ and $U=3.0$.
\begin{figure}
\caption{The N\'eel temperature, $T_{\rm N}$, 
  determined self-consistently as a function of $|t_1|$ 
 at $U=3.0$, and $\a_{\rm N}=0.992$ (square) and $\a_{\rm N}=0.994$ (circle).
An inset shows the $T_{\rm N}$  using $G^0$ in $\chi$ at $U=1.8$. 
The incommensurate region corresponds to the closed symbols.}
\label{phasediagram}
\end{figure}

In this model,  the nesting vector $\Q$ is commensurate, or ($\pi$,$\pi$),
 except  in the large $|t_1|$ and small $T$ region.
In Fig.~\ref{phasediagram}
the incommensurate region corresponds to the closed circles and squares 
 for $T\simle 0.02$.
The nesting vector is incommensurate in the $k_y$ direction,
but on the other hand it is always commensurate in the $k_x$ direction
         whereas the RPA result is not.\cite{Hasegawa1986,nestingVectorComment}
In other words,
the SDW phase  is locked to $\pi$ in the $k_x$ direction.
This result is consistent with the experimental nesting vector\cite{Takahashi1986,Nakamura1995}
 and is first realized in the self-consistent calculation.
We note that 
the nesting vector depends 
 on the details of the Fermi surface. 
Therefore it is necessary to use a realistic Fermi surface 
  in order  to compare the calculated nesting vector 
     with the experimental one for all  regions of the phase diagram.

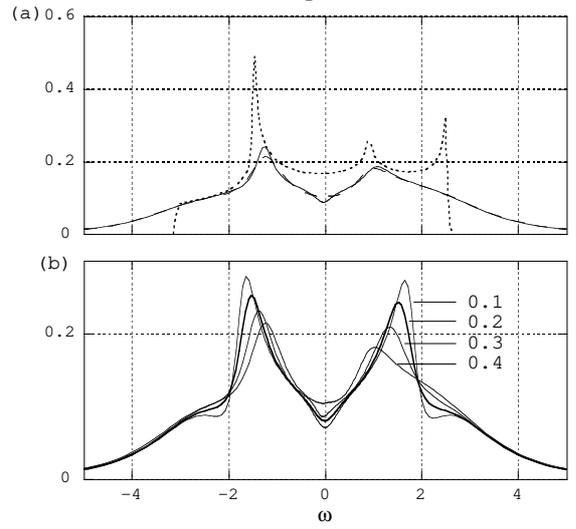
\begin{figure}
\caption{Fermi surfaces at $U=3.0$ (a solid line) and $U=0$ (a broken line) at $t_1=-0.4$ and $T=0.05$. }
\label{Fermisurface}
\end{figure}

Next let us see the Fermi surface.
The Fermi surface in the present model is flatter 
near $k_x \sim \pm \pi$ and pointed near $k_x \sim 0$ 
due to the $\cos(2bk_y)$ term. 
In Fig.~\ref{Fermisurface}, 
the Fermi surface is shown at $t_1=-0.4$, and $T=0.05$ and $U=3.0$.
The Fermi surface with interaction has a smaller band dispersion 
        in the $k_y$ direction,
at the same time,  the nesting condition is better than the one without interaction.

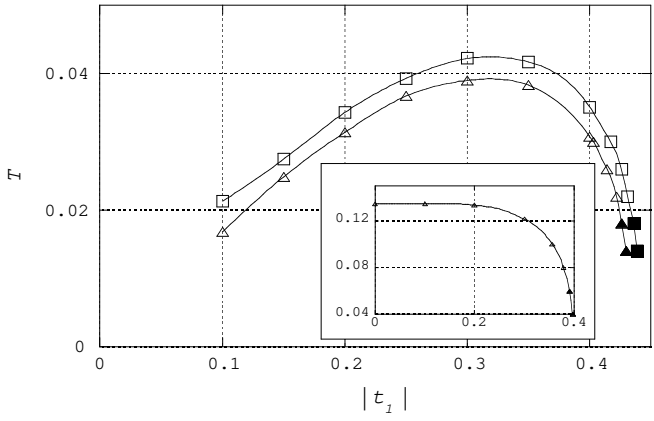
\begin{figure}
\caption{(a)The density of states at $t_1=-0.4$  and $U=3.0$, 
and  $T=0.05$ (a solid line) and $T=0.1$ (a broken line). 
and at $t_1=-0.4$ and $U=0$ (a dotted line).
(b)The density of states at $U=3.0$ and $T=0.1$ and 
$|t_1|=$0.1, 0.2, 0.3, and 0.4, respectively. }
\label{DOS}
\end{figure}

We plot the DOS in Fig.~\ref{DOS}.
The pseudogap emerges  near the chemical potential
 due to the spin fluctuations
 and 
it evolves with decreasing $|t_1|$ and $T$.
This result has close connection with the pseudogap
 near the chemical potential found 
 in the XPS and ARPES experiments. \cite{Dardel1993,Zwick1997}
We note that this model is effective near the chemical potential.

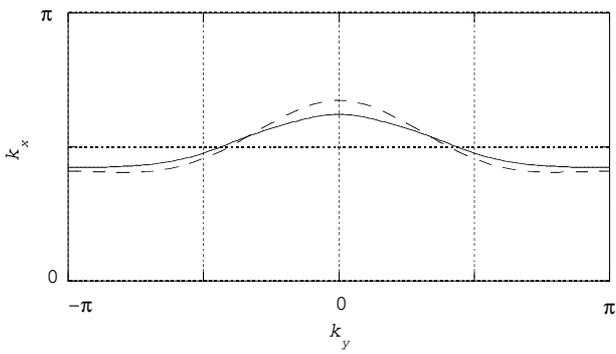
\begin{figure}
\caption{The temperature dependence of $\chi^{(-)}(0,\w=0)$ and  $1/T_1T$  at $t_1=-0.4$ and $U=3.0$.}
\label{susc}
\end{figure}

We also show  the uniform spin susceptibility ($\chi^{-}(0,\w=0)$) 
 and the NMR relaxation rate 
($(T_1T)^{-1} = \sum_k \lim_{\w\rightarrow 0}$ ${\rm Im}\;  \chi^{-}(k,\w)/\w$)
in Fig.~\ref{susc}.
With decreasing temperature, 
the uniform spin susceptibility  decreases gradually 
and this behavior reflects the evolution of the pseudogap in the DOS near the chemical potential. 
$(T_1T)^{-1}$ increases  with decreasing temperature 
in the low-temperature region.
Staggered susceptibility ($\chi^{-}(\Q,\w=0)$) is also the case.
These results seem to be consistent with  experimental results. \cite{Jerome1995,Wzietek1993}

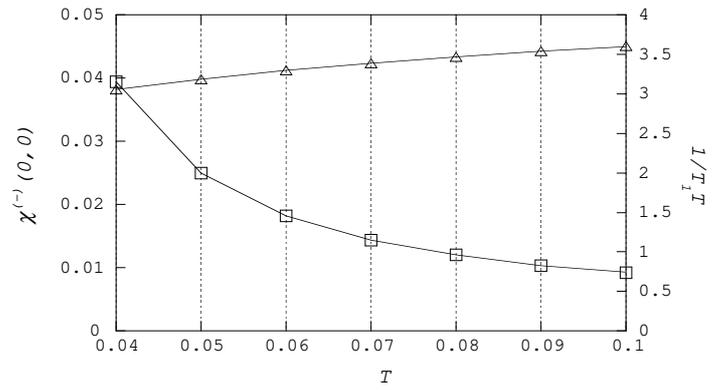
\begin{figure}
\caption{(a)The superconducting order parameter on the Fermi surface, 
$\phi(k_{\rm F},\w=0)$, 
from $k_y=-\pi$ to $k_y=\pi$ 
and (b)the Fermi surface and the contour plot of 
$\phi(\k,\w=0)$,
at $t_1=-0.45$, $U=3.0$ and $T=0.07$.}
\label{SCorderparameter}
\end{figure}

One can calculate the Green function at lower temperature than $T\sim 0.01$, 
though  the 256 Matsubara frequencies are not sufficient.
The superconducting condition is fulfilled  
at $t_1=-0.45$ and $T\simle 0.007$ in this model,
 when we use 128$\X$128 $\k$-mesh and 512 Matsubara frequencies.
We plot $\phi(k_{\rm F},0)$  corresponding 
to the singlet SC order parameter in Fig.~\ref{SCorderparameter}(a), 
where $T=0.007$ and $\l$ in eq.~(\ref{eqn:lambda}) equals 1.02.
One can clearly see that nodes exist  and
 a $d$-wave-like superconductivity is realized.
This is consistent with the NMR  experiment,
which predicts the SC order parameter with line-nodes.\cite{Takigawa1987}
There exists a dip  at $k_y\sim 0$ and a peak at $k_y \sim \pi$.
Figure~\ref{SCorderparameter}(b) shows 
a contour plot  of $\phi(\k,\w=0)$.
At $k_y=0$, the peak in $\phi(\k,\w=0)$ is closer
to $k_x=\pi/2$ than the Fermi surface,
and at $k_y=\pm \pi$ the situation is reversed.
These dip and peak structures depend
 on the manner of  warping of the Fermi surface and 
  disappear 
  when $|t_2|$ is small.
We have checked 
that $\lambda$ for the triplet case is 
about one third when the singlet SC is fulfilled. 
Therefore the triplet superconductivity is not realized in this model.


Below we focus on the obtained phase diagram in Fig.\ref{phasediagram}.
The self-consistent $T_{\rm N}$ declines with decreasing $|t_1|$,
whereas the non-self-consistent $T_{\rm N}$ does not descend with decreasing $|t_1|$.
At the same time,
the DOS at the chemical potential decreases with decreasing $|t_1|$
 because of the strong spin fluctuations 
in the self-consistent calculation as shown in Fig.~\ref{DOS}.
This effect is not taken into account when one uses $G^0$. 
These two behaviors are reasonably understood
through the Kramers-Kronig relation,
\begin{equation}
\chi^{(-)}(\k,\w=0) = \frac{2}{\pi}\; {\rm P}\! \int_{0}^{\infty}
        \frac{ {\rm Im} \chi^{(-)}(\k,\w+ i 0)}{\w} \, d\w
 \mbox{,}
\label{eq:KramersKronig}
\end{equation}
where the main contribution comes from  small $\w$.
The integrand  can be written as 
\( 1/\w \; {\rm Im}\chi^{(-)}(\k,\w) \simeq 
1/\w ({\rm Im} \chi^0(\k,\w))(1- U \chi^0(\k,\w) )^{-1}\)
 for large $\a_{\rm S}$, 
 and  
\(   1/\w \; {\rm Im}\chi^0(\k,+i0) \)  at zero temperature
is estimated as 
\begin{equation}
\lim_{\w\rightarrow 0} \frac{1}{\w} 
 {\rm Im}\chi^0(\k,+i0) = \sum_q {\rm Im} G(\q,0) \; {\rm Im} G(\q+\k,0) 
\mbox{.}
 \label{eq:ChiAndImG} 
\end{equation}
The right-hand side of eq.~(\ref{eq:ChiAndImG}) is roughly proportional to
 the square of the DOS at the Fermi level.
Therefore $ \chi^{(-)}(\k,0)$ decreases together 
 with the evolution of the pseudogap 
  in the DOS near the Fermi level.
This is the main reason why $T_{\rm N}$ decreases 
        for small $|t_1|$ in Fig.~\ref{phasediagram}.

One can interpret the reduction of $T_{\rm N}$ 
for small $|t_1|$ in  Fig.~\ref{phasediagram}
 in connection with the $S=1/2$ coupled chain 
 Heisenberg model.\cite{Miyazaki1995} 
The magnetization at $T=0$ decreases due to the zero-point fluctuation
 in decreasing  the interchain coupling.

On the other hand, when $|t_1|$ is large enough, 
then $|t_2|$ is also large, and 
 the nesting condition itself determines $T_{\rm N}$.
 Hence $T_{\rm N}$ drops rapidly as seen
 in the RPA and mean field calculations.

The effect of applied pressure can be considered
to increase the transverse transfer integrals  and the band width.
But it is sufficient only to enlarge the transverse transfer integrals 
in order to discuss the overall phase diagram in TMTTF/TMTSF salts.
Experimentally, the N\'eel temperature (or SDW transition temperature)
increases with increasing pressure, 
whereas it decreases and finally drops rapidly 
near the phase boundary with the superconducting phase.\cite{Moser1998}
We have derived, for the first time,
the phase diagram which coincides with the experimental one
 by considering the effects of the strong spin fluctuations self-consistently.

Next we compare the present results quantatively with 
 the experimental results.
If  $|t_0|$ is set to 1000 K, $T=0.01$ correspond to 10 K. 
These results do not correspond well with 
 the experimental ones
 because $U$ and $|t_2|$ are rather large
 in the present study due to the numerical calculations. 
But $T_{\rm N}$  is reduced with decreasing $U$ and 
 with increasing $|t_2|$,
 and 
the value of $t_1$ at which $T_{\rm N}$ takes its maximum also behaves 
        in a similar way.
Then one can adjust $U$ and   $|t_2|$ 
such that $T_{\rm N}$ and  $t_1$ are close to the experimental ones.


We have  calculated a quasi-one-dimensional model whose lattice has
zigzag hopping parameters, 
and have shown that the results are almost the same as that of the current model.
We have also calculated the quarter-filled Hubbard model without dimerization.
The temperature dependence  of the Stoner factor 
as a function of $t_1$
 is very similar to the case of the half-filled one in Fig.~\ref{phasediagram},
i.e., $T_{\rm N}$ has a peak as a function of $t_1$.
But 
 the resulting maximum $T_{\rm N}$ seems to be less than about $5\X 10^{-3}$
and the dimerization takes an important role quantitatively.
These results will be published elsewhere.

The recent experiments revealed 
that the antiferromagnetic phases were accompanied 
by  charge disproportionations.\cite{Pouget1996}
Naturally the importance of long-range Coulomb interactions was pointed out.\cite{Koyabashi1998,Seo1997}
The corresponding degrees of freedom, 
        which are dropped in the current model at half-filling,
 will not be negligible because of the weak dimerization.
Further studies including this effect are necessary 
 in connection with the nesting vector.


In summary, 
we have studied the model Hamiltonian 
 of the quasi-one-dimensional organic conductors by the FLEX method.
We have successfully explained the pressure dependence of the N\'eel temperature as a result of the evolution of the pseudogap in the DOS
        near the chemical potential,
 and the NMR relaxation rate which increases on cooling in the low-temperature region.
We have also found a  $d$-wave-like superconductivity 
 with line-nodes on the Fermi surface.

We are grateful to  T. Takahashi, F. Aryasetiawan, N. Katoh and J. Kishine for useful discussions.
We also thank Y. Kumagai for programming assistance.
This work is partly supported by NEDO.


\begin{thebibliography}{99}
\bibitem{Ishiguro1989}
T. Ishiguro and K. Yamaji: {\it Organic Superconductors} (Springer-Verlag).
\bibitem{Moser1998}
e.g.  J. Moser, {\it et al.}: Euro. Phys. J. B {\bf 1} 39-46 (1998).
\bibitem{Dardel1993}
B. Dardel, {\it et al.}: Europhys. Lett. {\bf 24} 687 (1993).
\bibitem{Zwick1997}
F. Zwick, {\it et al.}: Phys. Rev. Lett. {\bf 79} 3982 (1997).
\bibitem{Yamaji1983}
K. Yamaji: J. Phys. Soc. Jpn. {\bf 52} 1361 (1983).
\bibitem{Hasegawa1986}
Y. Hasegawa and H. Fukuyama: J. Phys. Soc. Jpn. {\bf 55} 3978 (1987).
\bibitem{Shimahara1989}
H. Shimahara: J. Phys. Soc. Jpn. {\bf 58} 1735 (1989).
\bibitem{Kuroki1998}
K. Kuroki and H. Aoki: preprint (cond-mat/9812026).
\bibitem{Kishine1998}
J. Kishine and K. Yonemitsu: J. Phys. Soc. Jpn. {\bf 67} 2590 (1998)
\bibitem{Grant1983}
P.M. Grant: J. Phys.  Colloque (France), {\bf 44} C3-847 (1983).
\bibitem{Jacobsen1983}
C.S. Jacobsen, D.B. Tanner, and K. Bechgaard: Phys. Rev.  B {\bf 28} 7019 (1983).
\bibitem{Penc1994}
K. Penc and F. Mila: Phys. Rev. B {\bf 50} 11429 (1994).
\bibitem{Bickers1991}
N.E. Bickers and S.R. White: Phys. Rev. B {\bf 43} 8044 (1991).
\bibitem{Bickers1989}
N.E. Bickers, D.J. Scalapino and S.R. White: Phys. Rev. Lett. {\bf 62} 961 (1989).
\bibitem{Kontani1998}
H. Kontani and K. Ueda: Phys. Rev. Lett. {\bf 80} 5619 (1998).
\bibitem{Kino1998} 
H. Kino and H. Kontani: J. Phys. Soc. Jpn. {\bf 67} 3691 (1998).
\bibitem{Kuroki1998highTc}
K. Kuroki and H. Aoki: J. Phys. Rev. Lett. {\bf 67} 1533 (1998).
\bibitem{high-Tc-Dahm} T. Dahm and L. Tewordt:
Phys. Rev.  B {\bf  52} (1995) 1297.
\bibitem{nestingVectorComment}
Within the fineness of the $\k$-mesh.
\bibitem{Takahashi1986}
T. Takahashi, {\it et al.}: J. Phys. Soc. Jpn. {\bf 55} 1364 (1986).
\bibitem{Nakamura1995}
T. Nakamura, {\it et al.}: Synth. Met. {\bf 70} 1293 (1995).
\bibitem{Jerome1995}
D. J\'erome, {\it et al.}: Synth. Met. {\bf 70} 719 (1995).
\bibitem{Wzietek1993}
P.  Wzietek, {\it et al.}: J. Phys. I (France) {\bf 3} 171 (1993).
\bibitem{Takigawa1987}
M. Takigawa, H. Yasuoka, and G. Saito: J. Phys. Soc. Jpn. {\bf 56} 873 (1987).
\bibitem{Miyazaki1995} 
T. Miyazaki, D. Yoshioka, and M. Ogata: Phys. Rev. B {\bf 51} 2966 (1995).
\bibitem{Pouget1996}
J.P. Pouget and S. Ravy:  J. Phys. I (France), {\bf 6} 1501 (1996), 
J.P. Pouget and S. Ravy:  Synth. Met. {\bf 85} 1523 (1997).
\bibitem{Seo1997}
H. Seo and H. Fukuyama: J. Phys. Soc. Jpn. {\bf 66} 1249 (1997).
\bibitem{Koyabashi1998}
N. Kobayashi, M. Ogata and K. Yonemitsu: J. Phys. Soc. Jpn. {\bf 67} 1098 (1998).


\end{thebibliography}
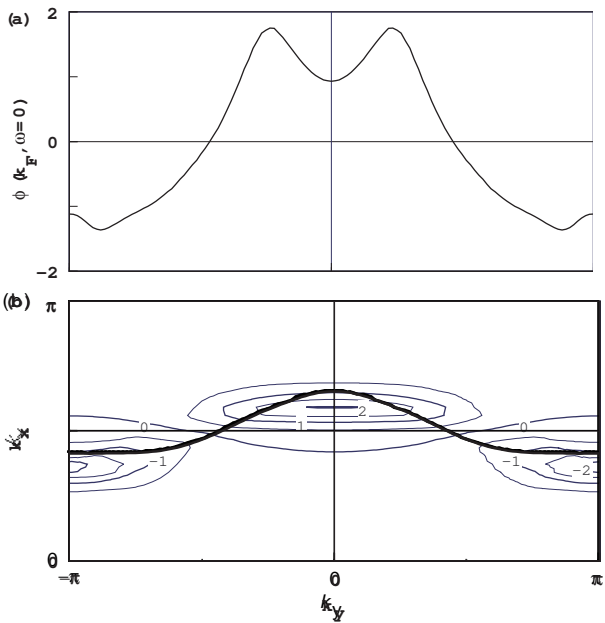
\end{document}